\documentclass{ws-ijbc}
\usepackage{ws-rotating}     
\usepackage{lscape}
\usepackage{color}

\begin{document}

\catchline{}{}{}{}{} 

\markboth{Pau Erola {\it et al.}}{Reliability of Optimal Linear Projection of Growing Scale-Free Networks}

\title{RELIABILITY OF OPTIMAL LINEAR PROJECTION OF GROWING SCALE-FREE NETWORKS}

\author{PAU EROLA}
\address{Department d'Enginyeria Inform\`{a}tica i Matem\`{a}tiques, Universitat Rovira i Virgili\\
Av. Pa\"isos Catalans 26 Tarragona, 43007 Catalonia, Spain\\
pau.erola@urv.cat}

\author{JAVIER BORGE-HOLTHOEFER}
\address{Instituto de Biocomputaci\'on y F\'\i sica de Sistemas Complejos (BIFI), Universidad de Zaragoza\\
Mariano Esquillor s/n 50018 Zaragoza, Spain\\
borge.holthoefer@gmail.com}

\author{SERGIO GOMEZ}
\address{Department d'Enginyeria Inform\`{a}tica i Matem\`{a}tiques, Universitat Rovira i Virgili\\
Av. Pa\"isos Catalans 26 Tarragona, 43007 Catalonia, Spain\\
sergio.gomez@urv.cat}

\author{ALEX ARENAS}
\address{Department d'Enginyeria Inform\`{a}tica i Matem\`{a}tiques, Universitat Rovira i Virgili\\
Av. Pa\"isos Catalans 26 Tarragona, 43007 Catalonia, Spain\\
alexandre.arenas@urv.cat}

\maketitle

\begin{history}
\received{(to be inserted by publisher)}
\end{history}

\begin{abstract}
Singular Value Decomposition (SVD) is a technique based on linear projection theory, which has been frequently used for data analysis. It constitutes an optimal (in the sense of least squares) decomposition of a matrix in the most relevant directions of the data variance. Usually, this information is used to reduce the dimensionality of the data set in a few principal projection directions, this is called Truncated Singular Value Decomposition (TSVD).
In situations where the data is continuously changing the projection might become obsolete. Since the change rate of data can be fast,
it is an interesting question whether the TSVD projection of the initial data is reliable. In the case of complex networks, this scenario is particularly important when considering network growth. Here we study the reliability of the TSVD projection of growing scale free networks, monitoring its evolution at global and local scales.
\end{abstract}



\section{Introduction}

There exists a vast literature that acknowledges Singular Value Decomposition as a valuable tool for information extraction from matrix-shaped data. This approach and its truncated variant have been extraordinarily successful in many applications \cite{golub96}, in particular for the analysis of relationships between a set of documents and the words they contain. In this case, the decomposition yields information about word-word, word-document and document-document semantic associations; the technique is known as latent semantic indexing \cite{berry1995using} (LSI) or latent semantic analysis \cite{landauer97} (LSA). In the field of complex networks, we recently introduced SVD as a useful tool to scrutinize the modular structure in networks \cite{arenas2010optimal}.

Remarkably, a common characteristic of these applications is their dynamic nature. In order to attain successful information retrieval, for instance in a query, LSI or LSA must rely on the fact that SVD of textual resources is always up to date. Unfortunately, databases rarely stay the same. Addition and/or removal of information is constant, meaning that catalogs and indexes quickly become obsolete or incomplete. Turning to networks, the question is equally pertinent: both natural and artificial networks are dynamic, in the sense that they change through time (and so do their modular structures). Paradigmatic examples of this fact are the Internet, the World Wide Web or knowledge databases like Wikipedia: all of them have been object of study from a graph-theoretical point of view \cite{pastor2004evolution,capocci2006preferential,zlatic06wiki}. Given this realistic scenario, a major question arise, namely, for how long TSVD stands as a reliable projection of evolving data.

In this paper we study the stability of TSVD as applied on changing networks. In particular, we want to quantify the differences between successive TSVD projections of evolving networks. To this end we devise a set of measures of global and local reliability, and apply them to a classical model of network growth, the Barab\'asi-Albert's (BA) scale-free network \cite{barabasi99}.
The BA model consist in a random network whose formation is driven by: growth, the network starts with a small number of nodes, and a new one is added at each time step; and preferential attachment, the probability of a new node $i$ linking to a previously existing node $j$ is proportional to the current degree of node $j$.
This mechanism yields networks with scale-free degree distributions $P(k) = k^{-\gamma}$.

This work is partially motivated by the application of TSVD to analyze the mesoscale of networks and its temporal evolution.
In \cite{arenas2010optimal}, the object of analysis is the \emph{contribution matrix} $C$, of $N$ nodes to $M$ modules. The rows of $C$ correspond to nodes, and the columns to modules. The analysis of this matrix is the focus of our research. The elements $C_{i\alpha}$ are the number of links that node $i$ dedicates to module $\alpha$, and is obtained as the matrix multiplication between the network's adjacency matrix $A$ and the {\em partition matrix} $S$:
\begin{equation}
C_{i\alpha} = \sum_{j=1}^N A_{ij} S_{j\alpha}\,,
\end{equation}
where $S_{j\alpha} = 1$ if node $j$ belongs to module $\alpha$, and $S_{j\alpha} = 0$ otherwise. Note that certain changes in the topology might not be reflected in the values of $C$, for example the rewiring of the connections of a node towards other nodes in the same community.

To measure the reliability of the TSVD projection of the contribution matrix, here we will  consider the ``worst case scenario" where each node belongs to its own community. This case corresponds mathematically to $C=A$. Establishing that TSVD is robust to change in these circumstances will settle the fact that TSVD is robust to change on a coarse-grained structure.




\section{Analysis of networks based on TSVD}
\label{svd}
Given a rectangular $N \times M$ (real or complex) matrix $A$, SVD stands for the factorization into the product of three other matrices,
\begin{equation}
  A=U \Sigma V^{\dag}\,,
\end{equation}
where $U$ is an unitary $N$-by-$N$ matrix (left singular vectors), and describes the original row entities as vectors of derived orthogonal factor values; $\Sigma$, the singular values, is a diagonal $N$-by-$M$ matrix containing scaling values; and $V^\dag$ denotes the conjugate transpose of $V$, an $M$-by-$M$ unitary matrix, which describes the original column entities in the same way as $U$.

A practical use of SVD is dimensional reduction approximation via truncation, TSVD. It consists in keeping only some of the largest singular values to produce a least squares optimal, lower rank order approximation. For example, severe dimensional reduction is a condition for success in machine learning SVD applications \cite{deerwester90,berry1995using,landauer98}.

In the case of a rank $r=2$ approximation, the unicity of the two-ranked decomposition is ensured if the ordered singular values $\sigma_i$ of the matrix $\Sigma$, satisfy $\sigma_1>\sigma_2>\sigma_3$ \cite{golub96}. This dimensional reduction is particularly interesting to depict results in a two-dimensional plot for visualization purposes.

The idea we developed in our previous work \cite{arenas2010optimal} is to compute the projection of the connectivity of nodes (rows in $A$) into the space spanned by the first two left singular vectors, we call this the projection space $\mathcal{U}_2$ and we denote the projected vector of the $i$-th node as $\tilde{n}_i$. Given that the transformation is information preserving \cite{chu2005inverse}, the map obtained gives an accurate representation of the main characteristics of the original data, visualizable and, in principle, easier to scrutinize. It is important to highlight that this approach has essential differences with classical pattern recognition techniques based on TSVD such as Principal Components Analysis (PCA) or, equivalently, Karhunen-Loeve expansions. Our data (columns of $A$) can not be independently shifted to mean zero without loosing its original meaning, this restriction prevents the straightforward application of the mentioned techniques.

\begin{figure}[!tpb]
  \begin{center}
  \begin{tabular}{l}
    {\bf a}
    \\
    \begin{tabular}[b]{l}
    \mbox{\includegraphics*[width=0.25\textwidth]{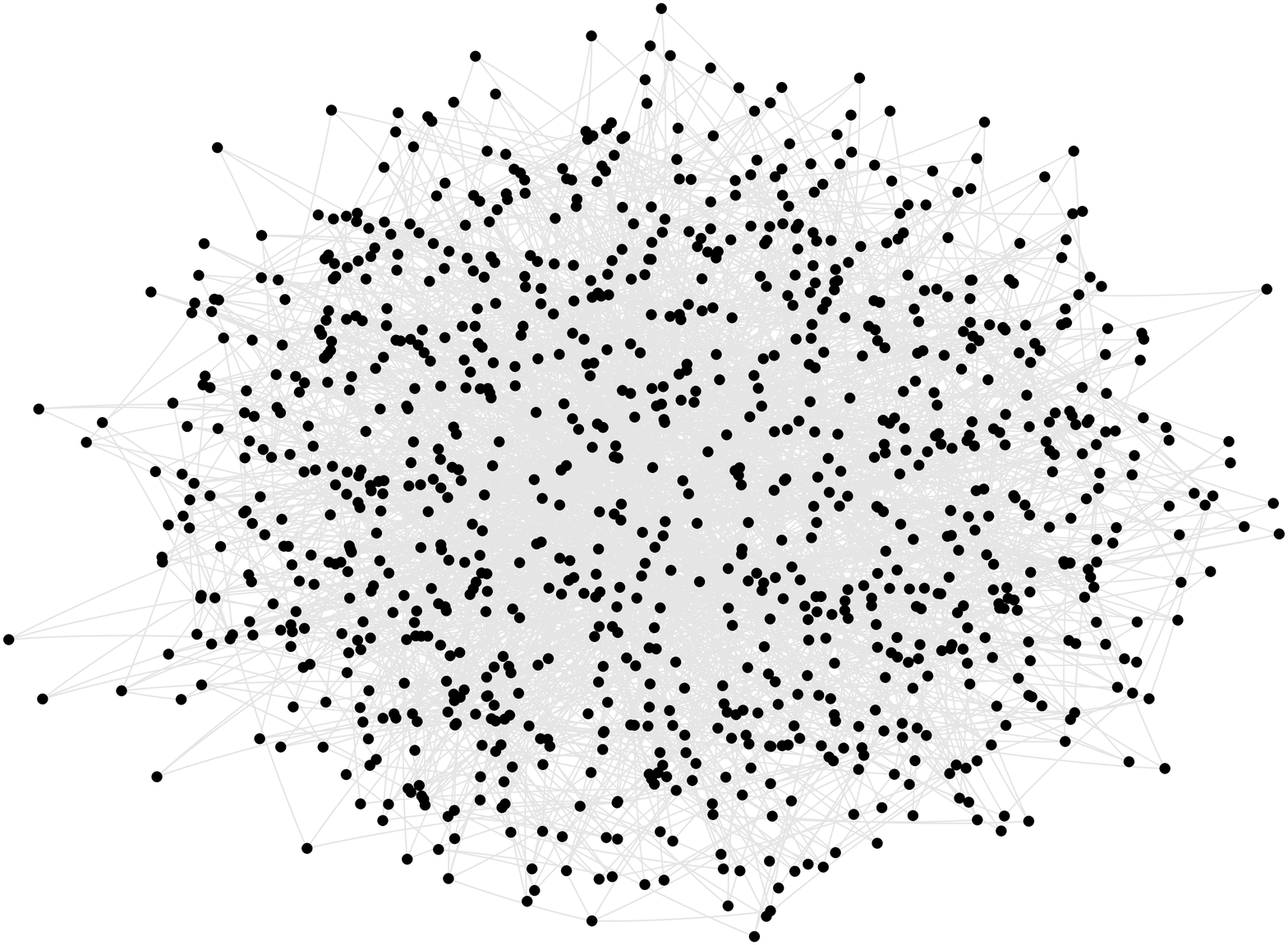}} \\ \mbox{\rule{0pt}{30pt}}
    \end{tabular}
    \mbox{\includegraphics*[width=0.42\textwidth]{sf1000rt.eps}}
    \begin{tabular}[b]{l}
    \mbox{\includegraphics*[width=0.25\textwidth]{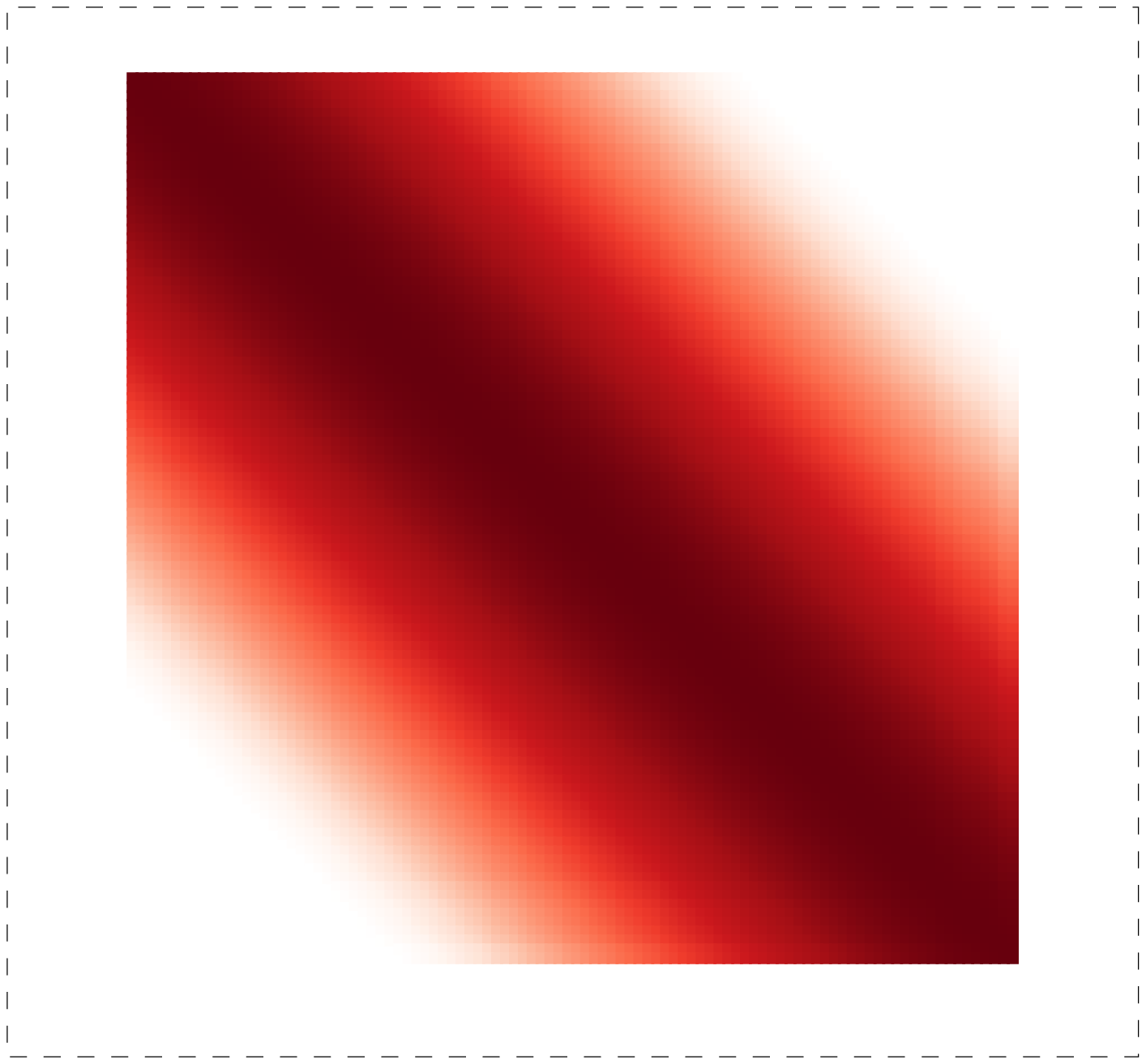}} \\ \mbox{\rule{0pt}{15pt}}
    \end{tabular}
    \\
    {\bf b}
    \\
    \begin{tabular}[b]{l}
    \mbox{\includegraphics*[width=0.25\textwidth]{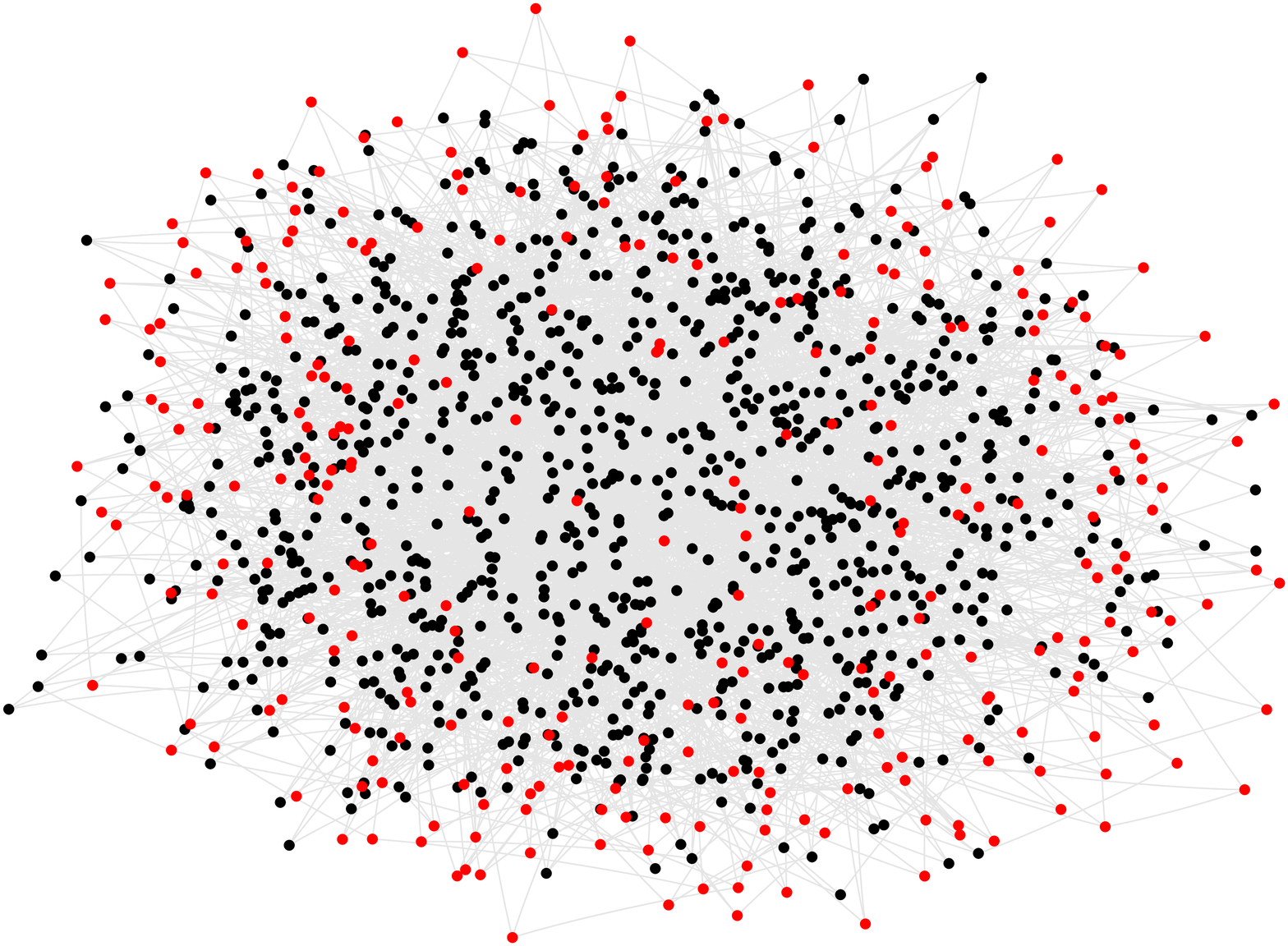}} \\ \mbox{\rule{0pt}{30pt}}
    \end{tabular}
    \mbox{\includegraphics*[width=0.42\textwidth]{sf1300rt.eps}}
    \begin{tabular}[b]{l}
    \mbox{\includegraphics*[width=0.25\textwidth]{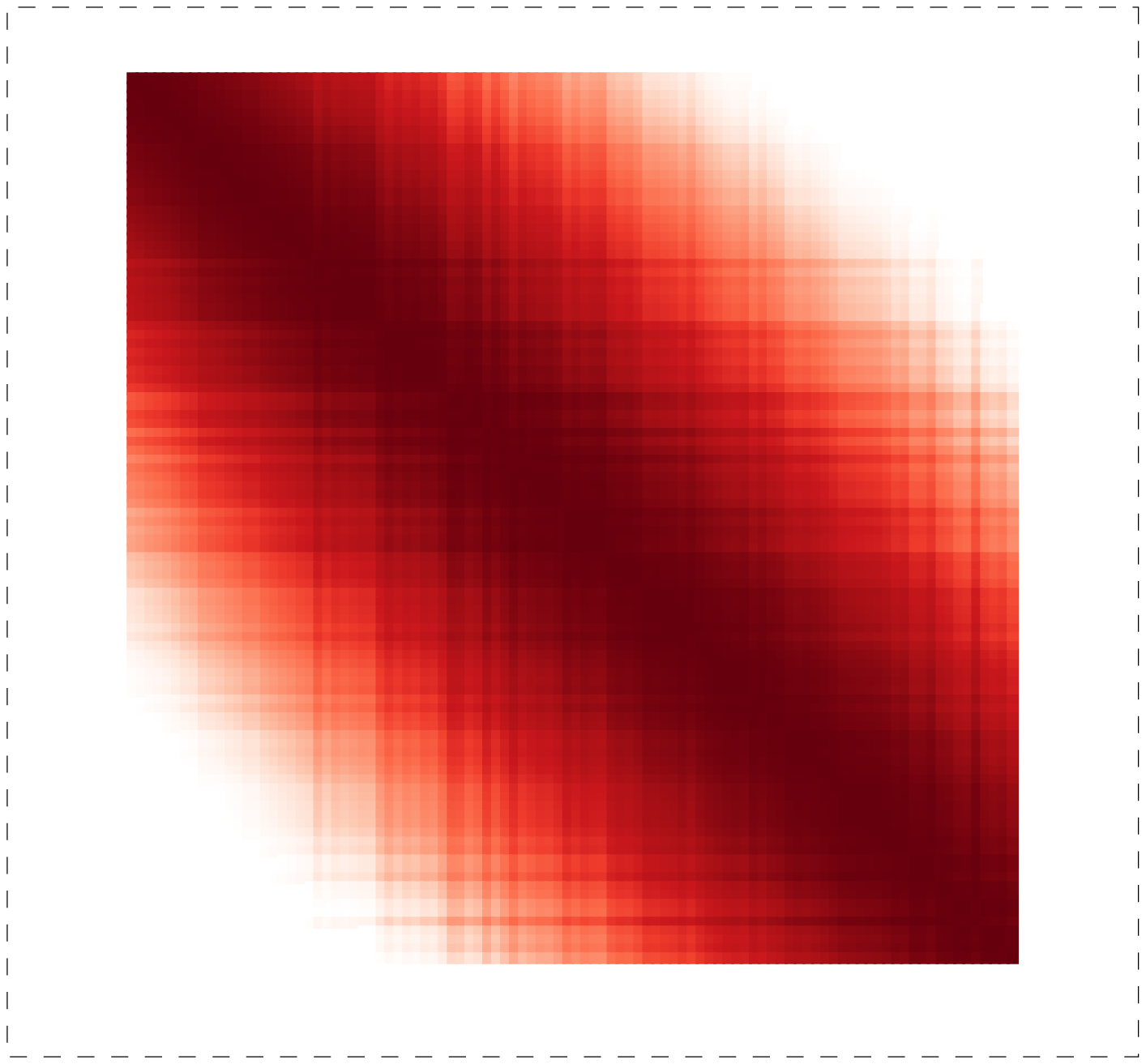}} \\ \mbox{\rule{0pt}{15pt}}
    \end{tabular}
    \\
    {\bf c}
    \\
    \begin{tabular}[b]{l}
    \mbox{\includegraphics*[width=0.25\textwidth]{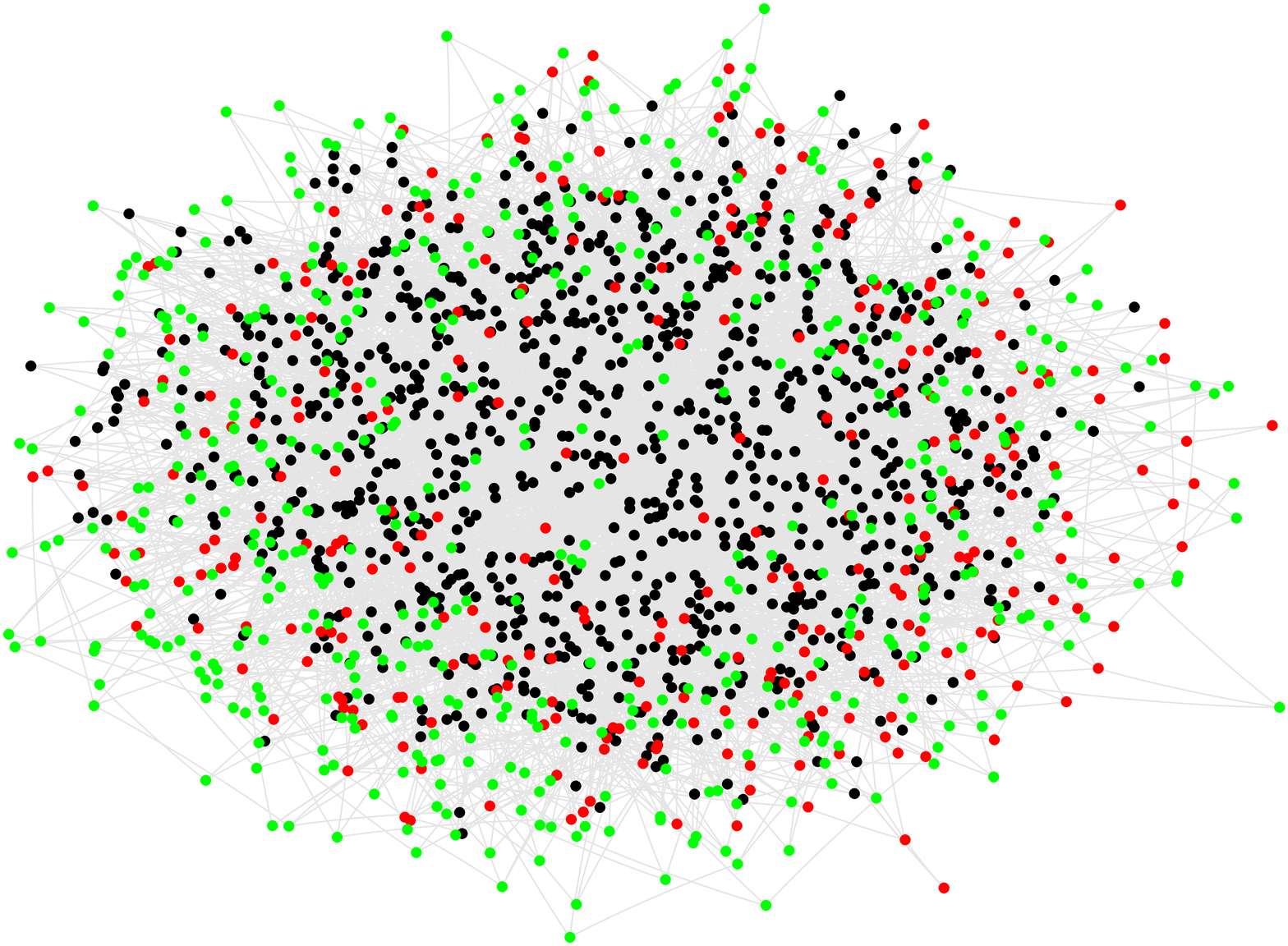}} \\ \mbox{\rule{0pt}{30pt}}
    \end{tabular}
    \mbox{\includegraphics*[width=0.42\textwidth]{sf1800rt.eps}}
    \begin{tabular}[b]{l}
    \mbox{\includegraphics*[width=0.25\textwidth]{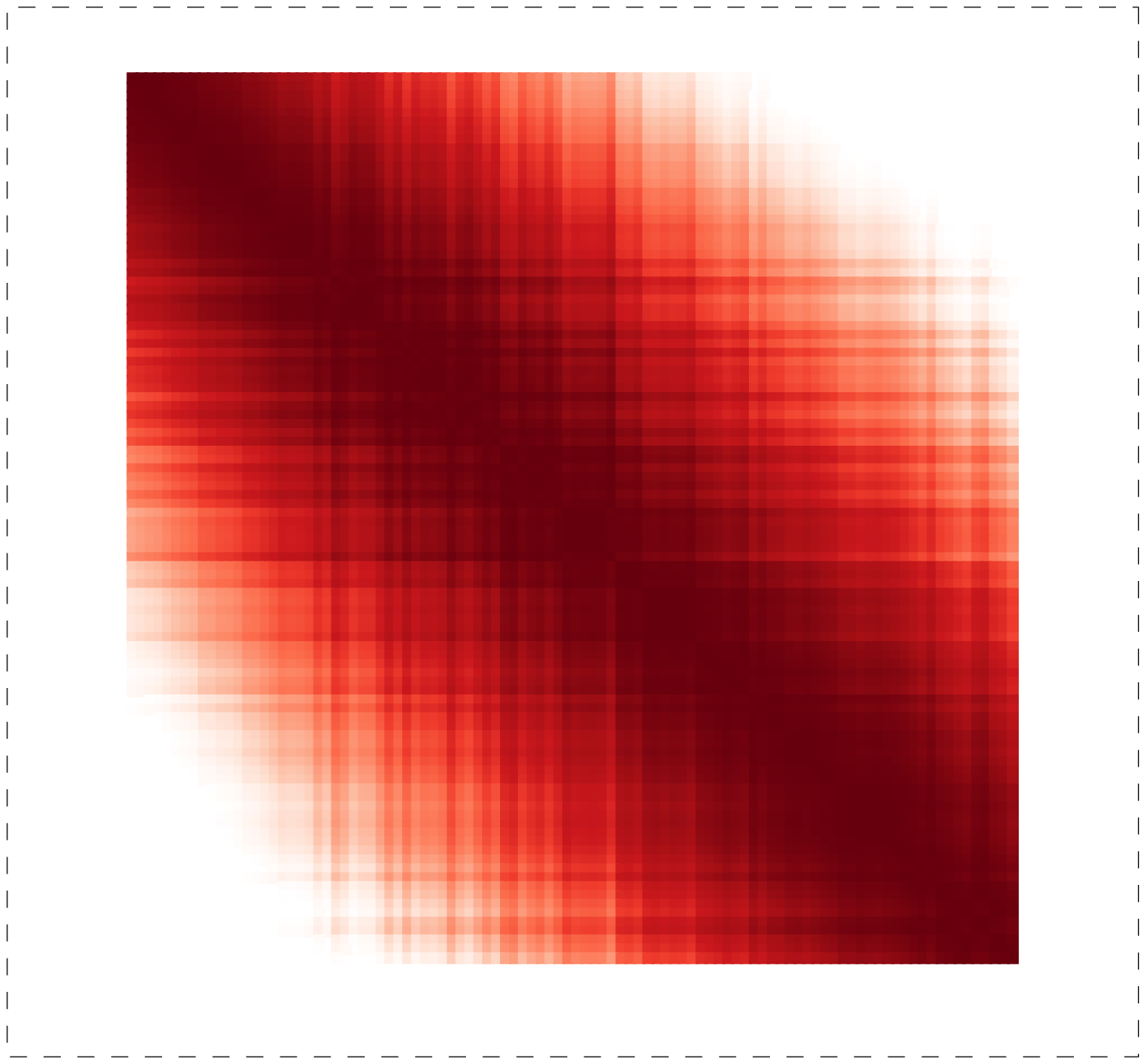}} \\ \mbox{\rule{0pt}{15pt}}
    \end{tabular}

  \end{tabular}
  \end{center}
\caption{Three snapshots of a growing network (left side), their corresponding projection on $R$--$\theta$ plane (center) and the $\theta$-overlapping matrices (right side; see \cite{arenas2010optimal} for details). For the sake of clarity, the initial set of nodes {\bf(a)} $N=1000$ are drawn in black; the second snapshot {\bf(b)} represents a growth of a 30\% of nodes, $N=1300$, new nodes are drawn in red. Finally {\bf(c)} represents a network with $N=1800$, last arrived nodes are depicted in green. Some nodes from the initial set have been highlighted (2, 4, 5, 6, 7, 964) in the $R$--$\theta$ plane, to get a visual intuition of the map's stability. Note that nodes with a high value of $R$ (2, 4, 5, 6, 7) remain almost unchanged throughout the topology's growth; whereas node 964 undergoes much change
from an absolute point of view. The rightmost matrices illustrate the amount of change of nodes with respect to their $\theta$ angles: as nodes are added in the structure the cosine overlaps between them increasingly distorts the original figure.
}
  \label{fig:rtmap}
\end{figure}

To interpret correctly the outcome of the TSVD we change to polar coordinates, where for each node $i$ the radius $R_i$ measures the length of its contribution projection vector $\tilde{n}_i$, and $\theta_i$ the angle between $\tilde{n}_i$ and the horizontal axis. Large values of $R$ correspond to highly connected nodes, and $\theta$ reflects the adjacencies of each node in matrix $A$. Fig.~\ref{fig:rtmap} shows the $R$--$\theta$ planes of an evolving network to get a visual intuition of the map's stability: as the network grows the mapping is distorted. In the following section we develop measures to quantify the effect of the growth on the TSVD projection.

\section{Quantifying the reliability of TSVD on growing networks}

As stated in the introduction, the goal of this research is to test how TSVD projection, at rank $r=2$, changes by computing it at different stages of the evolution of BA scale-free networks. This implies that TSVD will be computed on an initial network of size $N_0$, and then re-computed for successive node additions up to a final size $N_f = 2N_0$. To quantify the effect of growth on TSVD projection, we devise two levels of study: global and local. We will define measures based on the concept of absolute and relative distances between nodes, to this end we will work in the metric space $\mathcal{U}_2$.

\subsection{Global measure}

\noindent We propose a global quantity that indicates the amount of change in the position of nodes in the map obtained by TSVD. In the sequence of computed TSVD projections, the nodes' coordinates in $\mathcal{U}_2$ space change. This can be quantified by the difference of vectors $\tilde{n}_i$ between the initial and evolved network projection.

\begin{figure}[!tpb]
  \begin{center}
  \begin{tabular}{l}
    \mbox{\includegraphics*[width=0.5\textwidth]{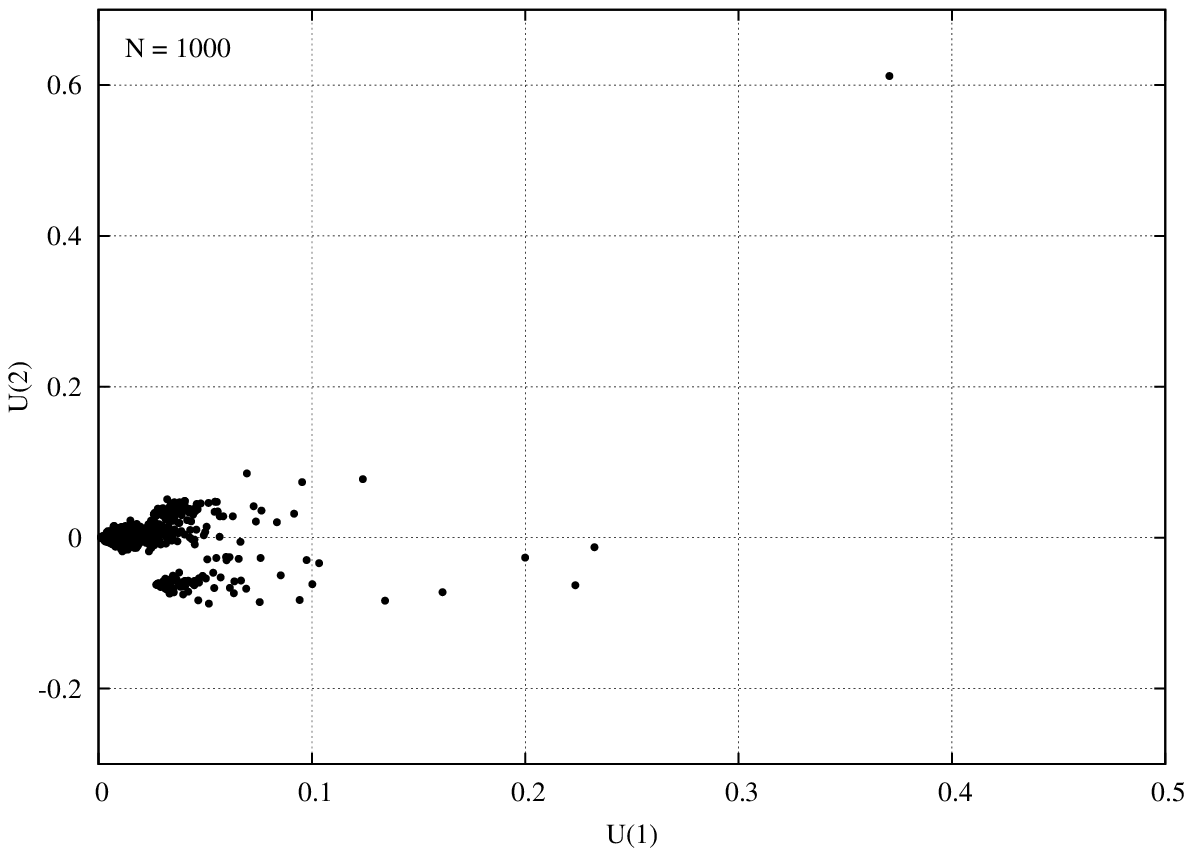}}
    \\
    \mbox{\includegraphics*[width=0.5\textwidth]{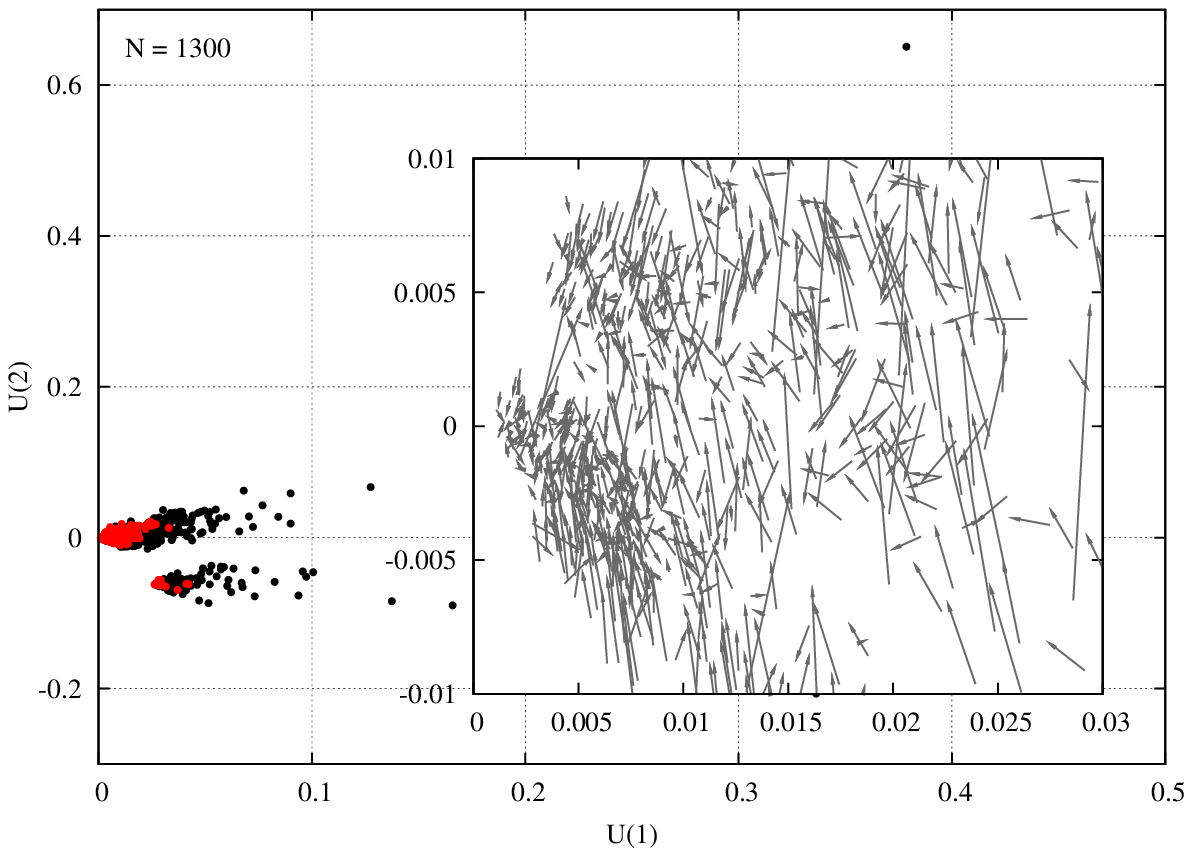}}
    \\
    \mbox{\includegraphics*[width=0.5\textwidth]{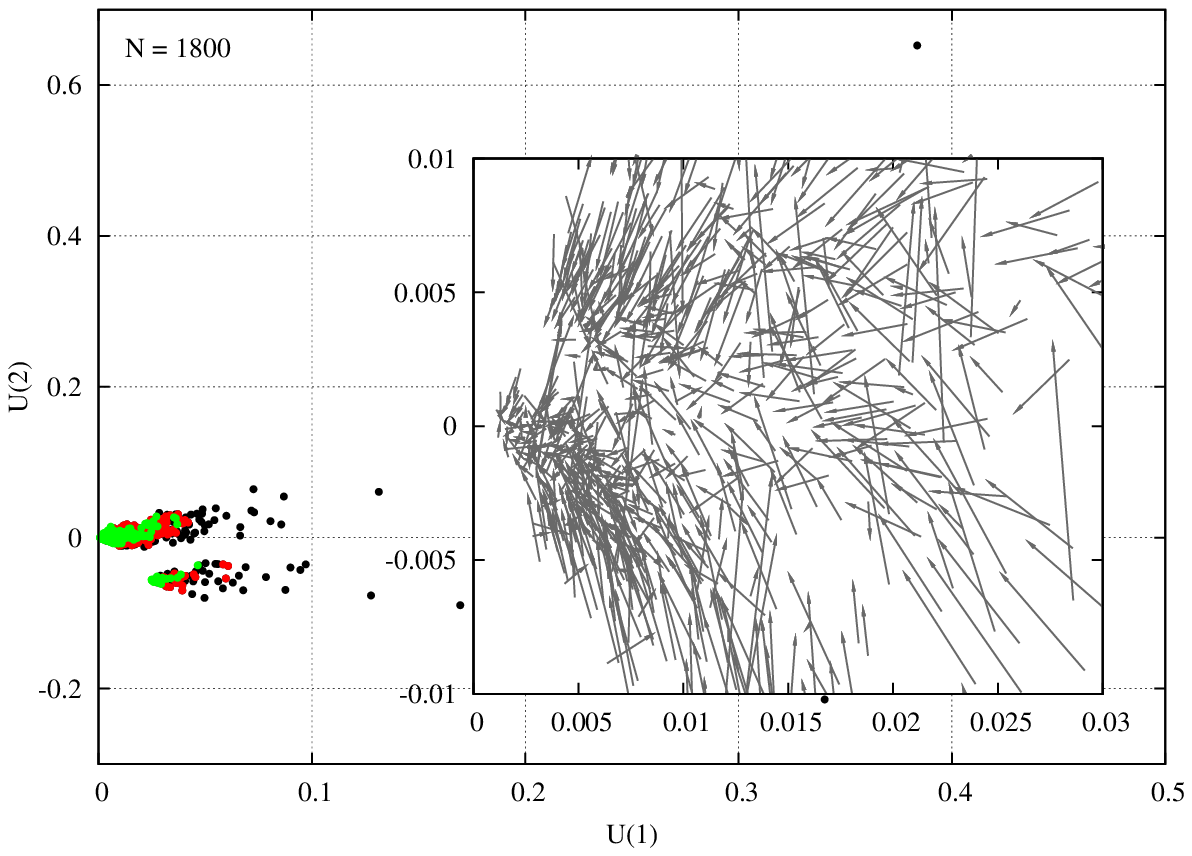}}
  \end{tabular}
  \end{center}
\caption{Projection in the $\mathcal{U}_2$ space of the evolving network presented in Fig.~\ref{fig:rtmap}. The insets for $N=1300$ and $N=1800$ trace the vectors $\tilde{v}_i$ between the projected coordinates of each node on the grown network and the original coordinates on the initial network with $N_0=1000$. We use these vectors $\tilde{v}_i$ to quantify the variability of the TSVD. Nodes are colored like in Fig.~\ref{fig:rtmap}. }
  \label{fig:vector}
\end{figure}

In Fig.~\ref{fig:vector} we plot the projection of the growing network presented in Fig.~\ref{fig:rtmap} on the space $\mathcal{U}_2$. We fix our attention in two time-shots of the evolution corresponding to growths of 30\% and 80\%. We compute the differences between positions of the same nodes at different stages ($z$) as $\tilde{v}_i= \tilde{n}^0_i - \tilde{n}^z_i$, producing a field map that accounts for the changes. This field map is shown in the insets of Fig.\ref{fig:vector}. When we have a 80\% increase of the initial size, the vectors $\tilde{v}_i$ are longer than in the 30\% increment, which evidences a larger variability, i.e. a progressive degradation in the TSVD reliability.

The global measure we propose to assess successive changes of rank $r$ TSVD projection compared to the initial data is computed by the {\em{relative error}}:

\begin{equation}
  E_{\mbox{\scriptsize global}} = \frac{\sum\limits^{N_0}_{i=1} \sum\limits^{r}_{j=1} | U^{z}_{ij} - U^{0}_{ij} |}{\sum\limits^{N_0}_{i=1}\sum\limits^{r}_{j=1}  | U^{0}_{ij} | }\,,
\end{equation}
where $U^0$ represents the truncated left singular vectors of the original network with $N_0$ nodes; and $U^z$ also represents the truncated left singular vectors, but of the grown network with size $N_z > N_0$.

\begin{figure}[!tpb]
  \begin{center}
    \mbox{\includegraphics*[width=0.5\textwidth]{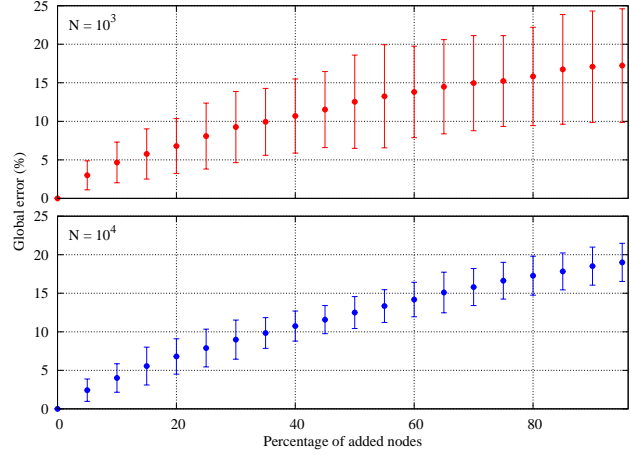}}
  \end{center}
\caption{Global error on two growing networks with initial sizes $N_0=1000$ (above) and $N_0=10000$ (below). For each network we compute the error by increments of 5\% of growth. In both cases, the global error is lower than 10\% up to the 40\% increment of network size. Each point is the average of 100 simulations.}
  \label{fig:global}
\end{figure}

We have applied this global measure to monitor the evolution of the TSVD stability for growing networks with initial sizes $N_{0}=1000$ and $N_{0}=10000$. Fig.~\ref{fig:global} shows the percentage of relative error with respect to the original network. In the chart, each successive point represents a 5\% of nodes addition. Up to a 40\% growth the global error remains below 10\%, and doubling the network size the average error still remains below 20\%. These results show the reliability of the projection after the growing process.


\subsection{Local measure}
Though informative, the previous global quantity can overlook changes at the microscopic level. The neighborhood of each node in the  $\mathcal{U}_2$ plane could undergo changes in the sequence of computed TSVD projections difficult to be revealed by the global measure defined above. Thus, we propose a measure that reflects these local changes using the distances between nodes in a neighborhood. Instead of defining a sharp border for the neighbors of each node, we propose to use a gaussian neighborhood that weights the distances according to a variance $\sigma$.

\begin{figure}[!tpb]
  \begin{center}
  \begin{tabular}{l}
    \mbox{\includegraphics*[width=0.5\textwidth]{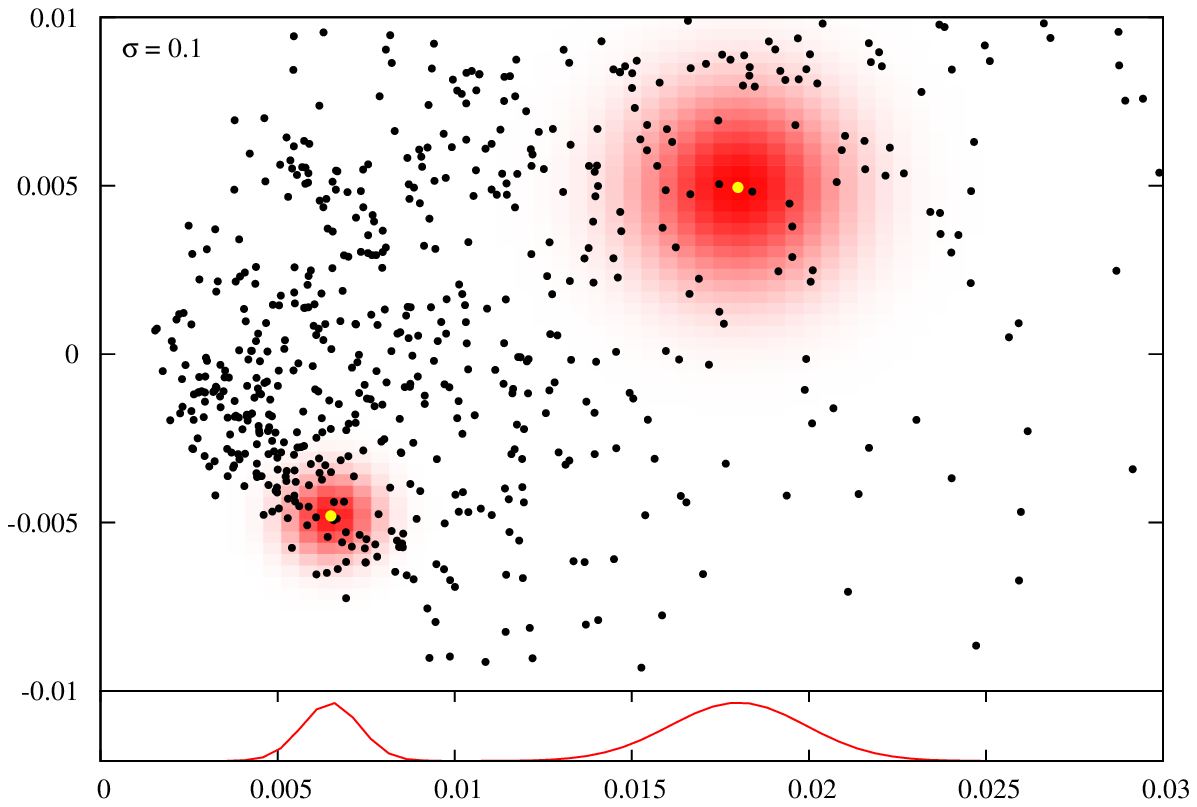}}
    \\
    \mbox{\includegraphics*[width=0.5\textwidth]{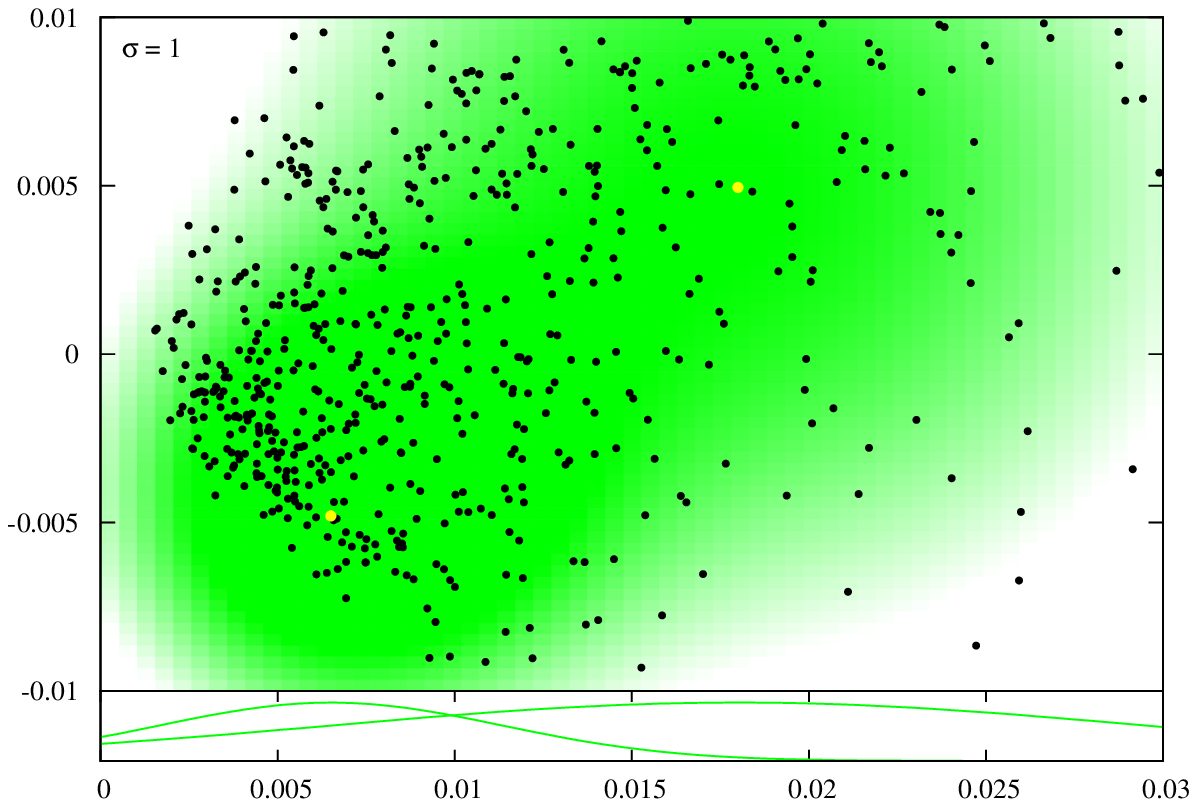}}
    \\
    \mbox{\includegraphics*[width=0.5\textwidth]{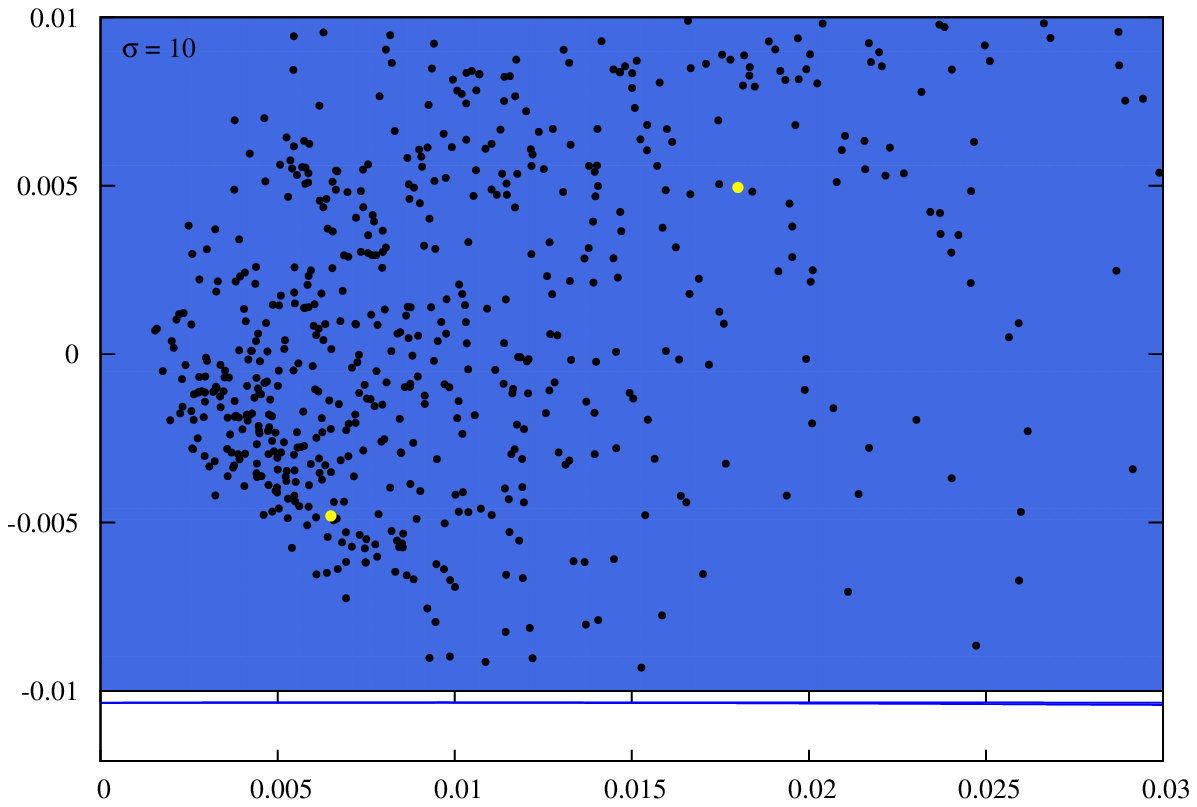}}
  \end{tabular}
  \end{center}
\caption{From top to bottom, we present the radius of influence of $\sigma=0.1$ (red), $\sigma=1$ (green) and $\sigma=10$ (blue). In the bottom of each chart, we have plotted, for the nodes highlighted in yellow, the gaussian curves that we added to matrix $D^z$ to compute the matrix of weighted distances $S^z$.
}
  \label{fig:gauss}
\end{figure}

First, we construct the $N \times N$ matrix of distances between any pair of nodes in the network at stage $z$ as
\begin{equation}
  D^z_{ij} = \sqrt{ \sum\limits^{r}_{k=1}  { \left( U^{z}_{ik} - U^{z}_{jk} \right) }^2 }\,,
\end{equation}
where $U^z$ represents the truncated left singular vectors of the network. These distances reflect a measure of proximity between nodes, independently on the global positioning in the map. The neighborhood is weighted to prioritize the stability on closer nodes over the distant ones. To this respect, we compute a matrix of weighted distances $S^z$ using a gaussian distribution that establishes a radius of influence as follows:

\begin{equation}
  S^z_{ij} = D^z_{ij} e^{ - \frac{ {D^{z}_{ij}}^2}{ 2{\left(  \sigma R^{0}_{i} \right)}^2}}\,,
\end{equation}
where we have chosen a radius of influence depending on the node. $R^{0}_{i}$ is the module of the projected vector $\tilde{n}_i$ in the initial network, and $\sigma$ is a constant.
This radius of influence proportional to the distance to the origin, emphasizes nodes with larger $R$ which are the most connected ones, see \cite{arenas2010optimal}. Using different values of $\sigma$ in the gaussian function we can tune the size of the neighborhood. Fig.~\ref{fig:gauss} shows, for a network with 1000 nodes, three magnified views of a network projection in $\mathcal{U}_2$ to illustrate the gaussian radius of influence.

Finally, the local measure of reliability we propose is computed as the {\em{relative error}}:
\begin{equation}
  E_{\mbox{\scriptsize local}} = \frac{\sum\limits^{N_0}_{i=1} \sum\limits^{N_0}_{j=1} | S^{z}_{ij} - S^{0}_{ij} |}{\sum\limits^{N_0}_{i=1}\sum\limits^{N_0}_{j=1} | S^{0}_{ij} |}\,,
\end{equation}
where $S^0$ and $S^z$ represent the matrices of weighted distances of the original network with $N_0$ nodes, and the grown network with size $N_z > N_0$, respectively.

\begin{figure}[!tpb]
  \begin{center}
  \begin{tabular}{l}

    \mbox{\includegraphics*[width=0.5\textwidth]{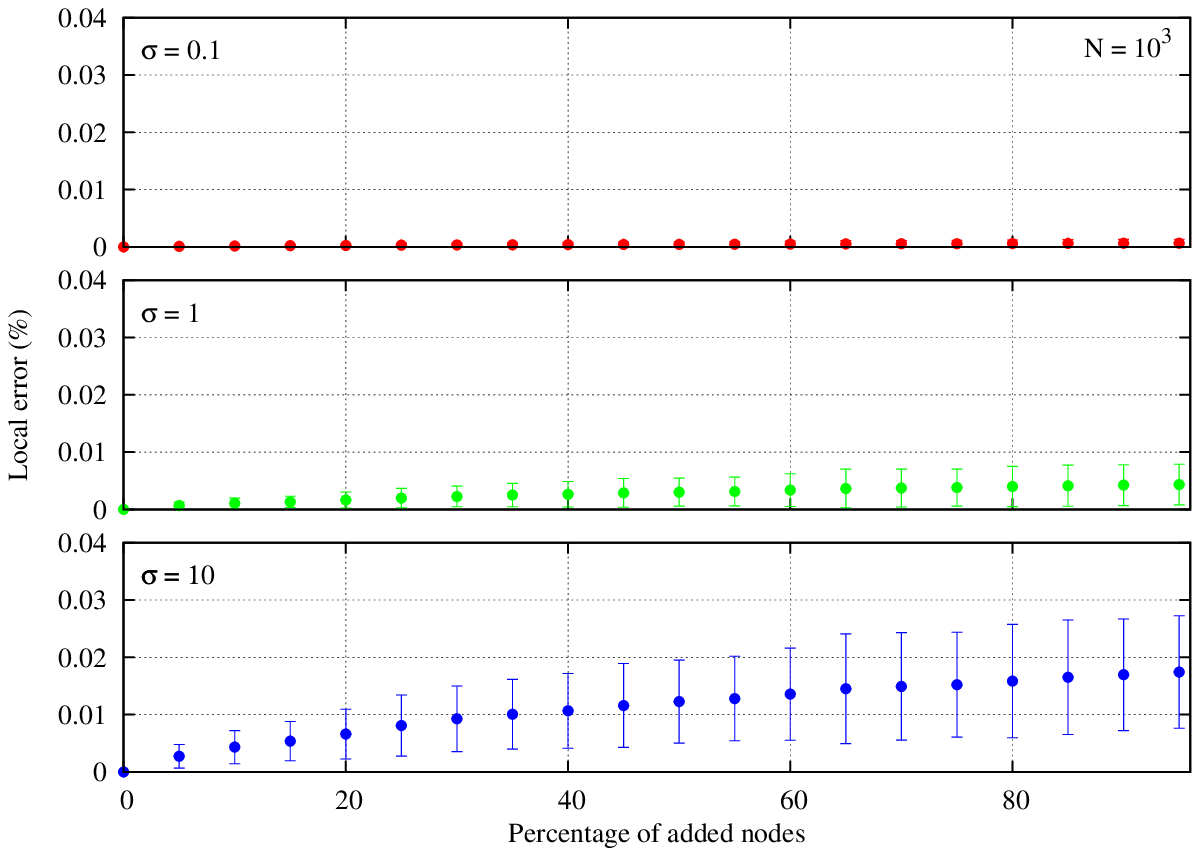}}
    \mbox{\includegraphics*[width=0.5\textwidth]{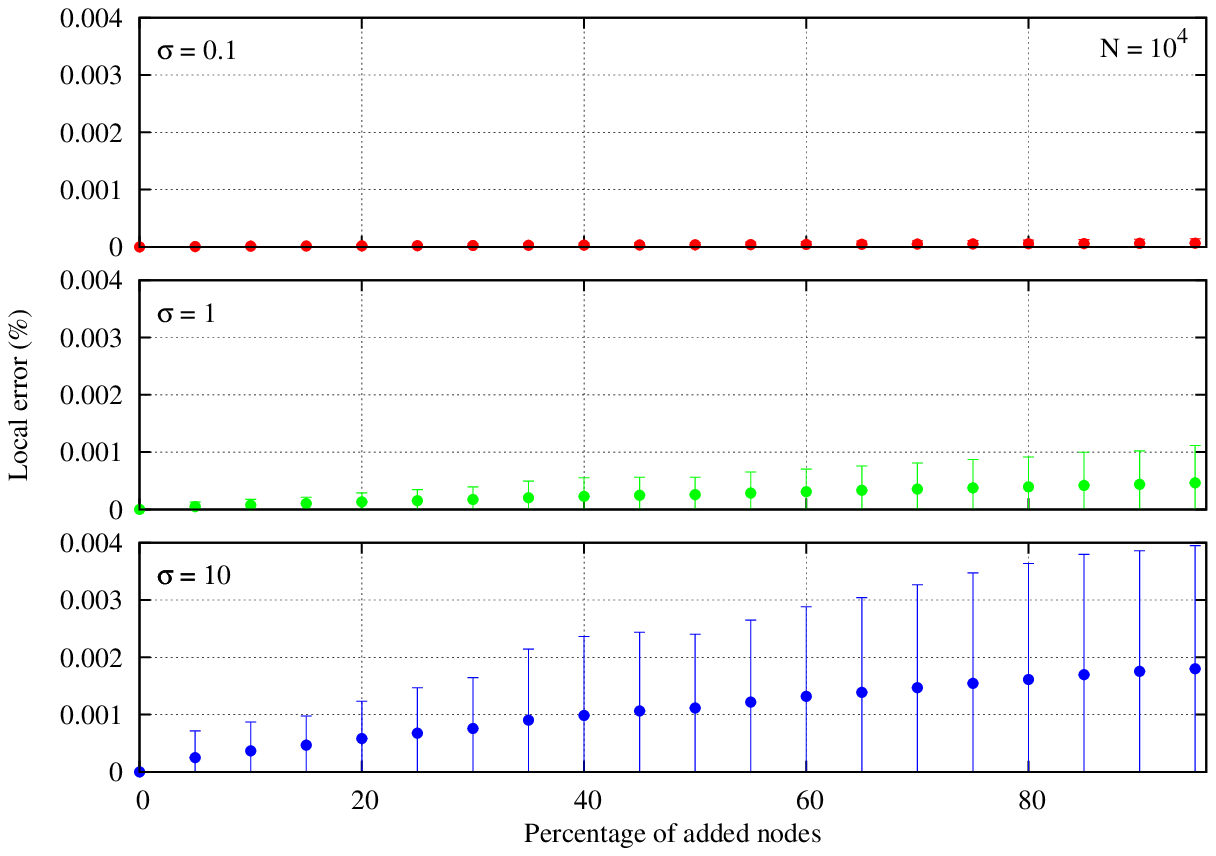}}

  \end{tabular}
  \end{center}
\caption{Local error on two growing networks with initial size $N_0=1000$ (left) and $N_0=10000$ (right). For each network we compute the relative error for $\sigma =$ $0.1$, $1$ and $10$ by increments of 5\% of growth. For small values of $\sigma$ the error is lower, but in all settings, the reliability of the projection is high. Each point is calculated with 100 simulations.}
  \label{fig:local}

\end{figure}

Fig.~\ref{fig:local} shows the local error measured on two growing networks by increments of 5\% of growth. Their initial size is $N_0=1000$ (left) and $N_0=10000$ (right). For each network we compute the relative error for $\sigma =$ $0.1$, $1$ and $10$.
When $\sigma=0.1$ only the closest neighbors have a significant weight in the measurement of the local error. These low values of $\sigma$ give the neighbor-wise error a very local sense.
On the other hand, when $\sigma=10$ the gaussian curve becomes flat and the measure is affected by the entire network perturbations, i.e. every node is equally considered as belonging to the neighborhood. Despite this global neighborhood for high $\sigma$ values, the local error measure represents a relative distance to each node, and as we see, doubling the network size the average error remains below 0.1\%. These very low error rates ensure a good reliability of the projection from a local point of view.

\section{Conclusions}
In this article we have raised the question about reliability of a standard linear projection technique such as SVD. The question is pertinent because SVD, and in particular its truncated version (TSVD), is rooted at the heart of some methodologies which pretend to extract useful and reliable information from dynamic data, i.e. data that is constantly undergoing change. We focus on growing scale-free networks.

We tackle the problem from two complementary points of view. At the large-scale level, we monitor average changes in nodes' TSVD projections. This means that each node's projection is compared against itself on successive changes.

Note however that success in practical applications of TSVD depends mostly on neighborhood stability. In other words, coherence of the output when data has suffered changes relies on the fact that the surroundings of a projected node are similar to those before those changes had happened. From a mathematical point of view, this merely implies that projections change in a coordinated way, such that relative positions are stable.
Keeping this in mind, the local measure developed above captures this facet of the problem by comparing not the evolution of a nodes position against itself, but rather against the rest of nodes. Furthermore, we introduce a parameter to weight this variation depending on the distance from the node of interest. This tunable parameter allows for a finer observation of neighborhood stability, ranging from immediate neighborhood measures to far-reaching areas. Note that the local measure is orders of magnitude lower than the global one. This points to the fact that, although the projection changes significantly, displacements in the plane $\mathcal{U}_2$ are similar in magnitude and direction on average. In other words, as a node of the network grows following the preferential attachment, it is highly likely that its neighbors also increase their weight staying close together.

Results indicate that TSVD projections are very robust against data growth. From a global point of view, an addition of 40\% of new data implies only an average change of 10\% from initial conditions. Doubling the amount of nodes to a network supposes a modification of 15\% in the positions of the set of initial nodes. More importantly, changes at the local level (neighborhood) are close to 0 even in the most demanding case.

Such results have been obtained with rather large structures ($N_{0}=10^{3}$ and $N_{0}=10^{4}$), which at the end of the process have doubled their initial size. This ensures that TSVD is reliable in a wide range of situations. On the other hand, our study focuses on a particular network model (BA) in which time plays an important role: the later a node appears, the lowest its chances to become an important one (a hub). We anticipate that the irruption of important entities  at late stages of evolution would surely disrupt TSVD projections in a more significant way. Nonetheless, we stress that growing systems typically develop smoothly, so our conclusions can be safely held.

Finally, we can briefly relate these results to the original motivation of the manuscript, that is, a scenario where the modular structure of networks is taken into account. In that situation, the stability of a TSDV map in the case of network changes is granted given the above reported results. Then, the characterization of the role of nodes and modules in terms of SVD's output can be safely regarded as faithful even in the case of severe changes in the underlying topology.


\bibliographystyle{ws-ijbc}
\bibliography{ebga}

\begin{thebibliography}{11}
\newcommand{\enquote}[1]{``#1''}
\providecommand{\natexlab}[1]{#1}
\providecommand{\url}[1]{\texttt{#1}}
\providecommand{\urlprefix}{URL }
\expandafter\ifx\csname urlstyle\endcsname\relax
  \providecommand{\doi}[1]{doi:\discretionary{}{}{}#1}\else
  \providecommand{\doi}{doi:\discretionary{}{}{}\begingroup
  \urlstyle{rm}\Url}\fi

\bibitem[{Arenas \emph{et~al.}(2010)Arenas, Borge-Holthoefer, G\'{o}mez \&
  Zamora-L\'{o}pez}]{arenas2010optimal}
Arenas, A., Borge-Holthoefer, J., G\'{o}mez, S. \& Zamora-L\'{o}pez, G. [2010]
  \enquote{Optimal map of the modular structure of complex networks,} \emph{New
  Journal of Physics} \textbf{12},  053009.

\bibitem[{Barab\'{a}si \& Albert(1999)}]{barabasi99}
Barab\'{a}si, A. \& Albert, R. [1999] \enquote{Emergence of scaling in random
  networks,} \emph{Science} \textbf{286},  509.

\bibitem[{Berry \emph{et~al.}(1995)Berry, Dumais \& O'Brien}]{berry1995using}
Berry, M., Dumais, S. \& O'Brien, G. [1995] \enquote{Using linear algebra for
  intelligent information retrieval,} \emph{SIAM review} \textbf{37},
  573--595.

\bibitem[{Capocci \emph{et~al.}(2006)Capocci, Servedio, Colaiori, Buriol,
  Donato, Leonardi \& Caldarelli}]{capocci2006preferential}
Capocci, A., Servedio, V., Colaiori, F., Buriol, L., Donato, D., Leonardi, S.
  \& Caldarelli, G. [2006] \enquote{Preferential attachment in the growth of
  social networks: The internet encyclopedia wikipedia,} \emph{Physical Review
  E} \textbf{74},  36116.

\bibitem[{Chu \& Golub(2005)}]{chu2005inverse}
Chu, M. \& Golub, G. [2005] \emph{Inverse eigenvalue problems: theory,
  algorithms, and applications} (Oxford University Press, USA), ISBN
  0198566646.

\bibitem[{Deerwester \emph{et~al.}(1990)Deerwester, Dumais, Furnas, Landauer \&
  Harshman}]{deerwester90}
Deerwester, S., Dumais, S., Furnas, G., Landauer, T. \& Harshman, R. [1990]
  \enquote{Indexing by latent semantic analysis,} \emph{Journal of the American
  Society for Information Science} \textbf{41},  391--407.

\bibitem[{Golub \& Van~Loan(1996)}]{golub96}
Golub, G. \& Van~Loan, C. [1996] \emph{Matrix computations} (Johns Hopkins
  University Press, Baltimore, MD).

\bibitem[{Landauer \& Dumais(1997)}]{landauer97}
Landauer, T. \& Dumais, S. [1997] \enquote{A solution to {P}lato's problem: the
  {L}atent {S}emantic {A}nalysis theory of acquisition, induction, and
  representation of knowledge,} \emph{Psychol. Rev.} \textbf{104},  211--240.

\bibitem[{Landauer \emph{et~al.}(1998)Landauer, Foltz \& Laham}]{landauer98}
Landauer, T., Foltz, P.~W. \& Laham, D. [1998] \enquote{Introduction to
  {L}atent {S}emantic {A}nalysis,} \emph{Discourse Process} \textbf{25},
  259--284.

\bibitem[{Pastor-Satorras \& Vespignani(2004)}]{pastor2004evolution}
Pastor-Satorras, R. \& Vespignani, A. [2004] \emph{Evolution and structure of
  the Internet: A statistical physics approach} (Cambridge Univ Pr), ISBN
  0521826985.

\bibitem[{Zlati{\'c} \emph{et~al.}(2006)Zlati{\'c}, Bo{\v{z}}i{\v{c}}evi{\'c},
  {\v{S}}tefan{\v{c}}i{\'c} \& Domazet}]{zlatic06wiki}
Zlati{\'c}, V., Bo{\v{z}}i{\v{c}}evi{\'c}, M., {\v{S}}tefan{\v{c}}i{\'c}, H. \&
  Domazet, M. [2006] \enquote{Wikipedias: Collaborative web-based encyclopedias
  as complex networks,} \emph{Physical Review E} \textbf{74},  16115.

\end{thebibliography}

\end{document}